\documentstyle[twoside,fleqn,espcrc2,epsfig]{article}

\newcommand{\DD}{{\bf D}}
\newcommand{\mbold}[1]{\mbox{\boldmath $#1$}}

\newcommand{\imag}{{\rm i\hspace{0.13ex}}}

\newcommand{\bra}[1]{\langle #1 |}
\newcommand{\ket}[1]{| #1 \rangle}
\newcommand{\bgeq}{\begin{equation}}
\newcommand{\bgeqa}{\begin{eqnarray}}
\newcommand{\edeq}{\end{equation}}
\newcommand{\edeqa}{\end{eqnarray}}

\newcommand{\ainv}{a^{-1}}

\newcommand{\lqcd}{\Lambda_{\rm QCD}}

\newcommand{\PRD}[1]{Phys.\ Rev.\ \textbf{D{}#1}} 
\newcommand{\msbar}{\overline{\rm MS}}

\title{Spectrum and decay matrix elements of $B$ and $D$-mesons
in lattice NRQCD}

\author{Joachim Hein\address{Newman Laboratory of Nuclear Studies, 
Cornell University, Ithaca, NY 14853, USA}%
\thanks{Supported by the National Science Foundation}}

\begin{document}

\begin{abstract}
We discuss recent results on the excitation spectra of $B$ and
$D$-mesons obtained in the framework of non-relativistic lattice QCD
in the quenched approximation.  The results allow for the
determination of the $\msbar$-mass of
$m_{b,\msbar}(m_{b,\msbar})= 4.34(7)$~GeV in 
${\cal O}(\alpha_s^3)$ in the perturbative
matching. The determination of the decay constants $f_{B_s}$ and
$f_{D_s}$ is discussed in detail. Results for the matrix
elements of semi-leptonic $B$ to $D$ decays are shown.
\end{abstract}

\maketitle

\noindent\hspace*{-1.9mm}
\raisebox{6.5cm}[0ex][0ex]{
{\normalsize
\parbox{4cm}{
\textbf{\textsf{CLNS 00-1689}}\\
\textbf{\textsf{hep-ph/0009088}}}}
}\vspace*{-5ex}

\section{MOTIVATION}

In the standard model of elementary particle physics, $CP$-violation is
caused by a single phase in the CKM-matrix. At the present time it
is experimentally observed only within the Kaon system. Establishing
$CP$-violation also in the $B$-meson system is one of the major goals
for the $B$-factory experiments, such as the newly running BaBar,
Belle and CLEO-III experiments as well as the future hadron collider
experiment LHCb. Failure of all $CP$-violating processes to be described
by the single CKM-phase would provide direct evidence for \textit{new
physics} beyond the standard model.

The extraction of the elements of the CKM-matrix from the
above experiments is complicated due to the hadronic nature of the
initial and final states. Here accurate knowledge of hadronic
quantities, such as QCD form factors of the $B$-meson, is needed. Due
to confinement these quantities are genuine non-perturbative. Lattice
gauge theory provides a means to determine such properties of hadronic
states based on first principles in a model independent way. The
approximations made in present day calculations are expected to be
relaxed in future calculations.

\section{LATTICE NRQCD}
For the results presented here we use non-relativistic QCD (NRQCD)
\cite{thalep,LepD46} to formulate heavy $b$ and $c$-quarks on a
lattice with a lattice spacing $a$ which is not negligible against its
Compton wave length. For heavy-light mesons, this results in a
systematic expansion of the Hamiltonian in powers of the inverse heavy
quark mass $m_Q^{-1}$
\bgeq H = H_0 + \delta H + \delta H_{disc} \,.
\edeq
The leading kinetic term is given as
\bgeq H_0 = - \frac{\DD^2}{2m_Q}\,,
\edeq
where $\DD$ denotes a covariant lattice derivative. 
We also include the following relativistic corrections
\bgeqa
\lefteqn{\delta H =} \nonumber\\
&& 
- \frac{g}{2m_Q}{ \mbold\sigma \cdot  \mbold B}
+ \frac{\imag g}{8m_Q^2}( \DD \cdot  \mbold E - \mbold E \cdot  \DD)\nonumber\\
&&
- \frac{g}{8m_Q^2} \sigma\cdot ( \DD \times \mbold E -  \mbold E\times \DD)
-\frac{(\DD^2)^2}{8m_Q^3} \,.
\label{deltaheq}
\edeqa
To reduce the dependence on the lattice spacing, the
following discretisation corrections are also included
\bgeq
\delta H_{disc} =
a^2 \frac{\sum_i\DD^4_i}{24m_Q}
-a\frac{(\DD^2)^2}{16nm_Q^2}\,.
\edeq
The first term on the right-hand side corrects in spatial, the second
one in temporal direction. The $n$ denotes the stability parameter
used in the evolution equation to generate the heavy quark
propagators. The inclusion
of the improvement terms is important due to the
non-re\-normali\-sability of NRQCD. The action has to be improved at
finite values of the lattice spacing such
that the results become independent of $a$ within the achieved
accuracy. We did not include radiative corrections to the
prefactors of the individual terms of this expansion. However,
the dominant tadpole contributions to this coefficients are removed by
mean field theory \cite{tadpole}.

The light quarks are simulated with the clover action \cite{swaction}
and the standard plaquette action is used for the gluon
background. All the results reported here are calculated in the
quenched approximation. This means the neglection of 
vaccum polarisation effects due
to the light sea quarks. For some of the
results also partly unquenched results exist. These calculations include two
flavours of light sea quarks. This will be pointed out in the
individual subsections.

In the calculations the value of the lattice spacing is determined
from the mass of the $\rho$-meson. This is senseful, since the here
discussed $B$ and $D$-mesons have a typical momentum scale of $\lqcd$
as does the $\rho$. In the quenched theory the strong coupling
constant runs differently from nature. This procedure minimises the
quenching error compared to using a quantity of a different scale.

\section{MESON SPECTROSCOPY}
Due to heavy quark symmetry the excitation spectra of the $B$ and
$D$-mesons are highly related. The lowest lying states are
experimentally well understood \cite{pdg}, however the situation with
respect to radially and orbitally excited states is not as
satisfactory. In this situation we can learn about the accuracy of the
approximations done in the present calculation, as well as predict
masses for so far undiscovered states.

\subsection{Spectrum of the $B$}
The spectrum of the $B$-meson has been investigated in large detail in
\cite{arifa_spec,BDspec}. In summary these publications provide
results for three different values of the lattice spacing $a$. Within
the achieved accuracy, the individual results for the different
splittings agree with each other. This is particularly important because
of the above discussed non-renormalisability of NRQCD.
A subset of the investigated states are also reported by the JLQCD
Collaboration 
\cite{fbjlnrqcd} using similar methods. These results are in excellent
agreement.

\begin{figure}
\centerline{\epsfig{file=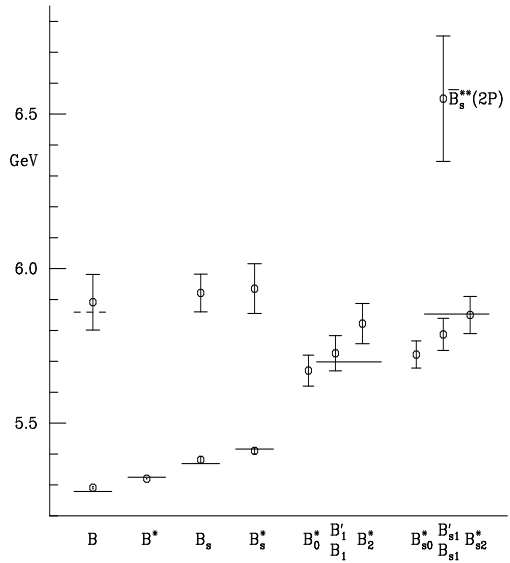,width=7cm,height=8cm}}
\caption{\label{Bspecfig} Excitation spectrum of the
 $B$-meson.}
\end{figure}

The results of \cite{arifa_spec,BDspec} are summarised in
figure~\ref{Bspecfig}.
The lattice results are displayed by the octagons, whereas
experimental results are given by horizontal lines.  We compare the
lattice $B^{**}$ and $B^{**}_s$ results to the experimental
$B^*_J(5732)$ and $B_{sJ}^{*}(5850)$ resonances.  The dashed line
gives a preliminary DELPHI result.  The overall agreement between
lattice and experiment is very good and the spin-independent spectrum
is nicely reproduced. This includes the radial and orbital excitation
energies as well as the strange non-strange $S$-wave splittings.
However there are problems with the
spin-dependent hyperfine splitting, which turns out to be
significantly too small. This might be due to the neglected radiative
corrections in eqn.~(\ref{deltaheq}), especially in front of the
$\mbold\sigma \cdot \mbold B$ term. Preliminary results \cite{c4pap}
indicate this might have an effect of the order of 10\% on the
hyperfine splitting. Another cause might be
the quenched approximation. Similar problems are also observed in
quenched light spectroscopy. The 
hyperfine splitting turns out increasingly too small 
with increasing quark mass \cite{yoshieed97}. 
So far results for the $B$-meson hyperfine splitting
with the inclusion of two flavours of light sea quarks do not show any
improvement for fixed lattice spacing
\cite{saradyna,arifapisa}. However these calculations use quite large
values for sea quark mass and improvement might only be seen once more
realistic values are used.

For the first time, this calculation gives a result for the radial
excited $P$-states, which is denoted as $B^{**}_s(2P)$ in the figure.

\subsection{Spectrum of the $D$}
NRQCD calculations of the $D$-meson within the here-presented
framework are only reliable for larger values of the lattice spacing.
The convergence of NRQCD for $D$-mesons has been investigated in
\cite{lewiswoloshyn}. A detailed analysis of the presented material
shows reasonable convergence of the NRQCD expansion for charmed heavy
light states \cite{BDspec}.

\begin{figure}
\centerline{\epsfig{file=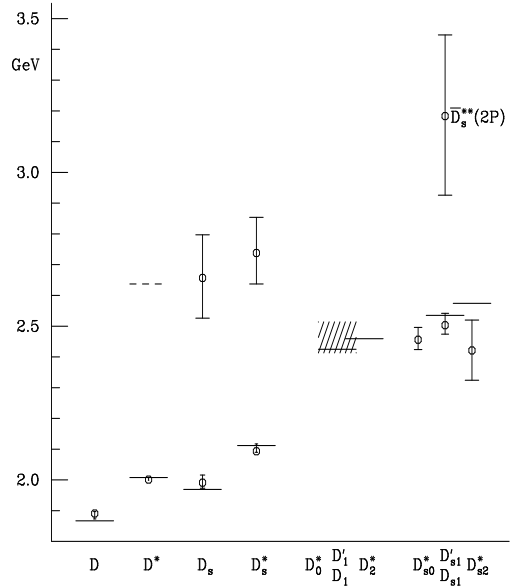,width=7cm,height=8cm}}
\caption{\label{Dspecfig} Excitation spectrum of the $D$-meson for
$\ainv = 1.116$ GeV.}
\end{figure}

The most complete results to date on the $D$-meson excitation spectrum
have been published in \cite{BDspec}. These results, using a lattice
spacing of 0.177 fm, are presented in figure \ref{Dspecfig}.
Again the symbols give the lattice results and the horizontal lines
the experimental outcome. The dashed line gives a result of the DELPHI
collaboration and the shaded region shows a preliminary result of the
CLEO collaboration for a wide $D_1$-resonance. Again the
spin-independent spectrum agrees very well with the experimental
outcome. Also the experimentally observed 
increase of the strange non-strange splitting from
the $B$ to the $D$-system is well reproduced. Again the hyperfine
splitting turns out to be too small however in this case by not as
much as for the $B$. This is consistent with the observed
mass-dependence of the hyperfine splitting for light quarks in the
quenched approximation \cite{yoshieed97}.  

Some of the splittings in figure \ref{Dspecfig} have also been
computed with heavy clover quarks \cite{peterhl,fermispec} and no
significant differences have been found between the heavy clover and
the NRQCD result. This is particularly remarkable for the $D^*_s-D_s$
hyperfine splitting, which strongly depends on the
details of the applied heavy quark action.  For charm quarks, NRQCD and
heavy clover quarks behave differently. The details
of this comparison are discussed in \cite{BDspec}.

\section{$\overline{\rm MS}$-MASS OF THE $b$-QUARK}
The mass of the $b$-quark is a fundamental parameter of the standard 
model. Based on the results of \cite{arifa_spec,BDspec} it is possible
to determine $m_b$ in NNNLO perturbation theory. Here preliminary
results of a forthcoming publication are presented \cite{mbmasscalc}.

\subsection{Calculation of the $b$-mass}
The calculation is done in two steps. First the pole mass is
calculated, which is converted into the $\msbar$-mass in a second
step.  The calculation of the pole mass turns out to be most precise
using 
\bgeq 
m_{b,{\rm pole}}= m_{\overline B} - \overline E_{\rm sim} + E_0\,.  
\edeq 
The mass $m_{\overline B}$ of the spin-average of the
$B$ and the $B^*$ is taken from experiment \cite{pdg} and the
simulation energy $\overline E_{\rm sim}$ from lattice
simulation. Here the use of the spin-averages eliminates the problems
encountered with the hyperfine splitting in the spectrum calculation.
The self-energy $E_0$ of the $b$-quark is known from perturbation
theory. The latter is available most precise in the limit of infinite
quark mass. It has been analytically calculated to ${\cal
O}(\alpha_s^2)$ \cite{martsach} and numerically to ${\cal
O}(\alpha_s^3)$ for the quenched case \cite{e0mont,direnzo}. The 
here-presented analysis is based on \cite{e0mont}. The simulation energies
of \cite{arifa_spec,BDspec} have been extrapolated to the static
limit.

\begin{figure}
\centerline{\epsfig{file=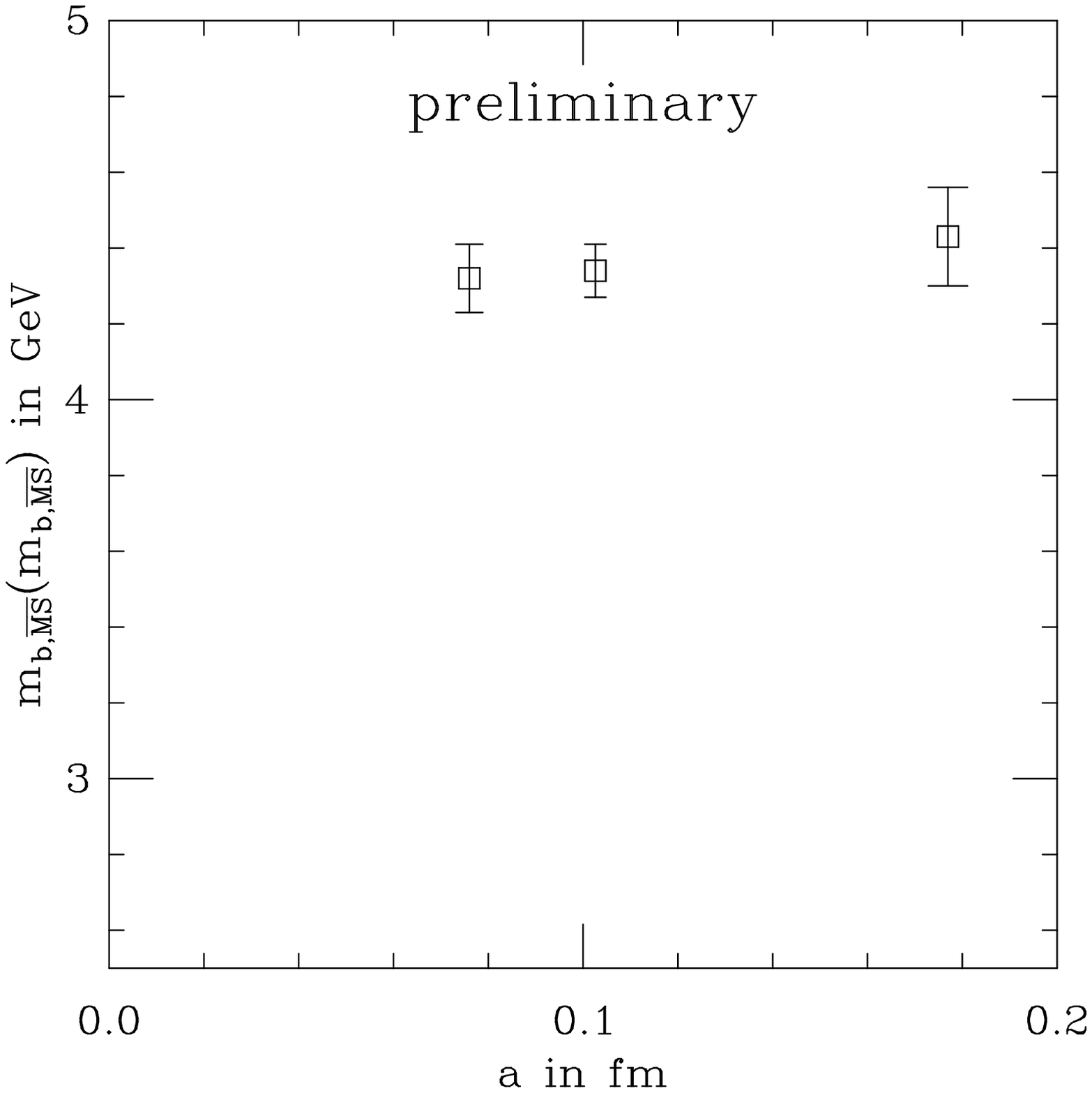,width=7cm}}
\caption{\label{mbscalefig} Dependence of $m_{b,\msbar}(m_{b,\msbar})$
on the lattice spacing $a$.}
\end{figure}

To convert the pole mass to the $\msbar$ scheme, we use the
renormalisation constant to ${\cal O}(\alpha_s^3)$ \cite{melnikov}.
The result for three different values of the lattice spacing is shown
in figure~\ref{mbscalefig}.
The different results agree within their statistical accuracy and
the most precise of them is quoted as the final result
\bgeq \label{mbresult}
m_{b,\msbar}(m_{b,\msbar}) = 4.34(7)\ {\rm GeV}\,.
\edeq
The uncertainty is still dominated by the uncertainties of $E_0$. It
further includes the uncertainties arising from the conversion to the
$\msbar$-scheme, statistics of $\overline E_{\rm sim}$, the lattice
spacing and the discretisation. Corrections due to the finite mass of
the $b$-quark are expected to be of ${\cal O}(\lqcd/m_b)$ of
$\overline E_{\rm sim}$, which is about 0.05 GeV. 

The effect of the quenched approximation can be estimated at ${\cal
O}(\alpha_s^2)$ \cite{martsach}, when using $\overline E_{\rm sim}$
with two flavours of light quarks from \cite{saradyna}. In comparison
to the quenched case in the same order of perturbation theory, the sea
quarks reduce
$m_{b,\msbar}(m_{b,\msbar})$ by 0.07 GeV. Eqn.~(\ref{mbresult}) is
compatible to the result in ${\cal O}(\alpha_s^2)$ of \cite{martsach}.

\section{LEPTONIC DECAY}
The pseudo-scalar decay 
constant $f_M$ determines the decay of a mesonic state $M$ 
into a pair
of leptons
\bgeq
p_\mu f_{M}= \langle 0 | A_\mu| M(p) \rangle\,.
\edeq
For the $B^+$-meson the leptonic width is highly CKM suppressed and 
is not expected to be measurable in the near future. Therefore the
decay constant cannot be measured directly, however it is an
important input parameter in other analysis, for example 
$B-\bar B$-mixing \cite{yamada}. The decay constant of the $B$ has to
be determined from theory. The $D_s$ has a much weaker CKM suppression and
its decay constant has been measured experimentally, which provides us with
the possibility to check the theoretical calculation.

When using NRQCD for the heavy quarks, the matrix element of the 
axial current of QCD has to
be expanded into matrix elements of the effective theory
\bgeqa \label{fbexpansion}
\langle A_0\rangle_{\rm QCD} &=& C_0 \langle \bar q \gamma_5 \gamma_0
Q\rangle\nonumber \\
&&- C_1 \frac{\langle\bar q \gamma_5 \gamma_0
(\mbold \gamma\cdot \DD) Q\rangle}{2m_Q} \nonumber\\
&&+ C_2 \frac{\langle(\mbold \DD\bar q \cdot \mbold\gamma) \gamma_5\gamma_0 Q\rangle}{2m_Q} \,.
\edeqa 
The coefficients $C_i$ on the right-hand side have been determined in
1-loop lattice perturbation theory \cite{colinjunko}.

\subsection{Investigations of $f_{B_s}$}
\begin{figure}
\centerline{\epsfig{file=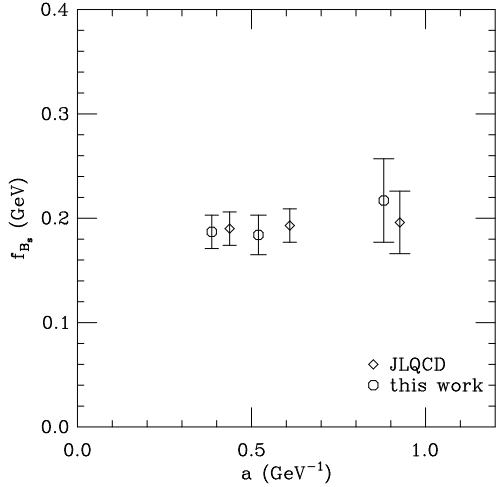,width=7cm}}
\caption{\label{fbsscalefig} Lattice spacing dependence of $f_{B_s}$.}
\end{figure}

The dependence of the result of $f_{B_s}$ on the lattice spacing is
one of the key issues of \cite{fbsfds}. The results have been combined
with those of \cite{arifafb} and are reproduced in
figure~\ref{fbsscalefig}. A similar investigation has been performed
in \cite{fbjlnrqcd}. These results are included in the figure as well.
The individual results agree well with each other and one can conclude
that the results are within the claimed accuracy 
indeed independent of the value of the lattice
spacing $a$. For the final number we quote
\bgeq \label{fbsresult}
f_{B_s} = 187 (16)~{\rm MeV}\,.
\edeq
A detailed breakdown of the error can be found in \cite{fbsfds}. 
The result
agrees well with the outcome of recent studies using
other ways to formulate the heavy quark on the lattice,
see \cite{draper,hashimoto} for
reviews. Calculations including two flavours of sea quarks indicate an increase
of approximately 30 MeV for $f_{B_s}$ \cite{saradyna,arifapisa}.

Reference \cite{fbsfds} investigates power law terms and the
$\lqcd/m_Q$ corrections of the decay constant in large detail.  Due to
the different dimensionality of the operators on the right hand side
of eqn.\ (\ref{fbexpansion}) the coefficient $C_0$ develops a $1/am_Q$
divergence, which cancels a similar divergence in the matrix
element $\langle\bar q \gamma_5 \gamma_0 (\mbold \gamma\cdot \DD)
Q\rangle$. The perturbative expansion of $C_0$ allows for the explicit
investigation of this cancellation in ${\cal O}(\alpha_s/am_Q)$.  The
remaining part of the ${\cal O}(1/m_Q)$ matrix elements in eqn.\
(\ref{fbexpansion}), which contains the ${\cal O}(\alpha_s^2/am_Q)$
part of the divergence as well as physical ${\cal O}(\lqcd/m_Q)$
terms, amounts to a few percent of the final result.

\begin{figure}
\centerline{\epsfig{file=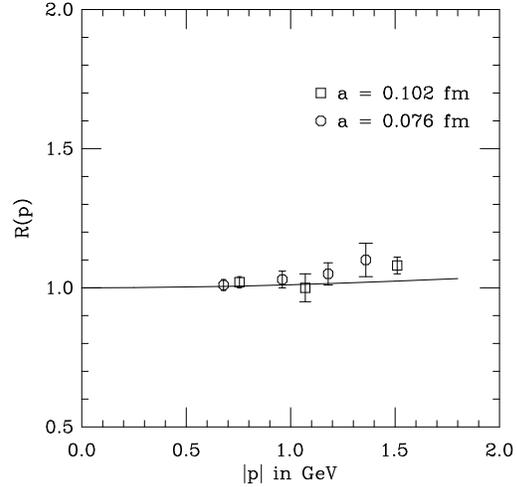,width=7cm}}
\caption{\label{fbmomfig} Investigation for 
momentum dependent discretisation effects on $f_{B_s}$.}
\end{figure}

The decay constant $f_{B_s}$ is a momentum independent form factor.
It allows the study of the effect the lattice discretisation has on the
form factor of moving mesons. This is an important consistency check
with respect to semi-leptonic decays. In this case the form factors
are momentum dependent, and discretisation effects and physical
effects are hard to disentangle. In figure~\ref{fbmomfig}  we
plot the ratio
\bgeq 
R(\vec p) = \frac{\bra{0}A_0\ket{B_s(\vec p)}/\sqrt{E(\vec p)}} 
{\bra{0}A_0\ket{B_s(0)}/\sqrt{E(0)}}\,.  
\edeq 
This has been investigated for two different values of $a$ \cite{fbsfds}.
In the absence of discretisation effects this becomes $\sqrt{E(\vec p)/E(0)}$, 
which is included as a full line.
The results are consitent with the continuum expectation up to momenta
of about 1.2 GeV. For the largest momenta the deviations are less than
$8\%$.

\subsection{Decay constant of the $D_s$}
The decay constant of the lighter $D_s$ was successfully determined in
lattice NRQCD \cite{fbsfds} as well. As for the spectrum, this was
done on coarse lattices. The result includes all the corrections up to
${\cal O}(\lqcd/m_Q)$. After cancellation of the power divergence in
${\cal O}(\alpha_s/am_c)$ the residual ${\cal O}(1/m_c)$ matrix
elements contribute only on the $10\%$ level to the final number,
indicating good convergence of the NRQCD expansion in the charm
region. The final outcome is \bgeq f_{D_s}= 223(54)~{\rm MeV}\,.
\edeq This result agrees well with the experimental result of
$f_{D_s}=280(19)(28)(34)$~MeV \cite{pdg2000} as well as other lattice
calculations in the quenched approximation \cite{draper,hashimoto}.

\section{SEMI-LEPTONIC DECAYS $B \to D l \nu$}\label{semlepsec}
Semi-leptonic decays of a $B$-meson into $D,D^*, D^{**},D',\dots$ provide the
best way to measure $|V_{cb}|$. A precise measurement of this is
required to relate $CP$-violation as measured from the $K$ to the one
measured from the $B$. In case of pseudo-scalar decay products such as
$D$ and $D'$, one needs to determine the matrix element
\bgeqa
\langle B|V_\mu|D\rangle \label{bdmatrixel}
& =&  \sqrt{m_Bm_D}[ h^+(\omega)(v_B\!+\!v_D)_\mu \nonumber \\
&&  + h^-(\omega)(v_B\!-\!v_D)_\mu ]\,.
\edeqa
As shown, this can be parametrised by two form factors $h^+(\omega)$
and $h^-(\omega)$ with $\omega=v_B \cdot v_D$.
In this section we summarise the results of \cite{BDproc}, which are
still preliminary.

\subsection{Elastic scattering}
\begin{figure}
\centerline{\epsfig{file=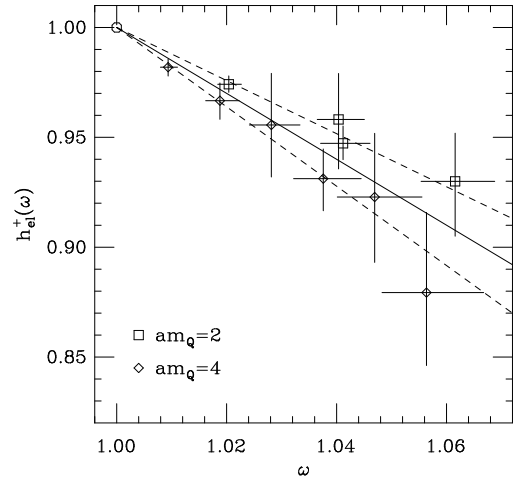,width=7cm}}
\caption{\label{iwfctfig} Form factor $h^+_{\rm el}$ in the case of 
elastic scattering as a function of the recoil parameter $\omega$.}
\end{figure}

The elastic scattering of a $B$-meson from a vector current provides
the simplest approximation to the matrix element in
eqn.~(\ref{bdmatrixel}). In the static limit, $m_B \to \infty$ this
becomes the Isgur-Wise function $\xi(\omega)$.  The
$\omega$-dependence of the elastic form factor has been studied for
two different values of the mass of the heavy quark. The result is
reproduced in figure~\ref{iwfctfig}.
The values for $am_Q=4$ corresponds approximately to a $B_s$-meson. 
The figure gives
\bgeq
\rho^2_{\rm strange}=1.5(3)(4)
\edeq
for the slope of the Isgur-Wise function in case of a strange
spectator quark.
The results have been checked for their dependence on the
momentum of the external states.

\subsection{Radial excited states}
\begin{figure}
\centerline{\epsfig{file=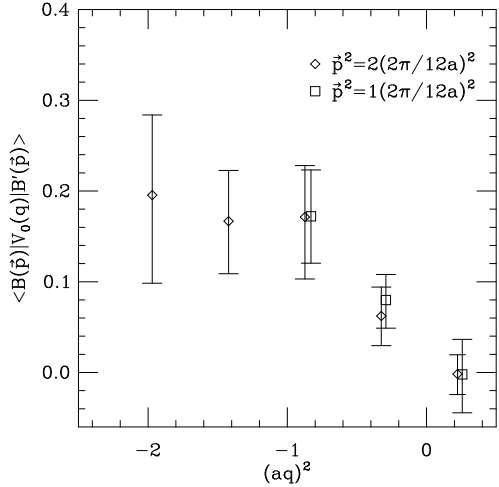,width=7cm}}
\caption{\label{radfffig} Form factor for transitions between
pseudoscalar ground and radially excited state.}
\end{figure}

The use of two different interpolating fields with the same quantum
numbers makes it possible to observe a signal for the matrix element
$\langle B_s |V_0(q)|B'_s \rangle$ involving a radial excited state.
The $q$ dependence of the matrix element is shown in
figure~\ref{radfffig}. 
For both external states the same heavy quark mass is used.
The external states are orthogonal at zero recoil and the matrix
element vanishes as expected. Once these states get boosted with
respect to each other, the matrix element becomes finite. This
result is qualitative and demonstrates the feasibility to determine
such matrix elements in lattice gauge theory.

\subsection{Non-massdegenerate transitions}
\begin{figure}
\centerline{\epsfig{file=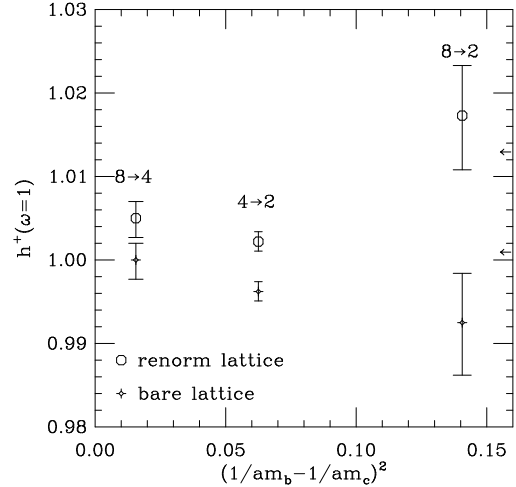,width=7cm}}
\caption{\label{h+omfig} Form factor $h^+$ at zero recoil.}
\end{figure}

The case of unequal heavy quark mass for the in and outgoing
state has been studied at zero recoil. In this case the vector current
is not a conserved current anymore and gets renormalised. The
renormalisation constants have been perturbatively calculated to ${\cal
O}(\alpha_s)$ \cite{peterpert}. Figure~\ref{h+omfig} gives the result
with and without this renormalisation for three different pairs of
heavy quark masses. 
The used heavy quark masses are indicated above
the individual results in units of the lattice spacing. Again the
value of $am_Q=4$ corresponds to the $B_s$-meson. The arrows at the
right hand side of the the figure give the upper and lower error bound
of the lattice result from \cite{fermiBD}. This has been extrapolated
to the physical $B\to D$ transition and corresponds to $(1/am_b -
1/am_c)^2 \approx 0.56$.

\section{DISCUSSION}
This talk summarises our recent results on the physics of heavy light
mesonic states. The results include the $B$ and $D$-meson spectrum as
well as the decay constants of the $B_s$ and the $D_s$. The results
for the $B$ and the $B_s$ have been obtained for three different values of
the lattice spacing and the results are found to be independent of the
lattice spacing within error bars. In the spectrum we observe good
agreement to experiment for spin independent splittings. However the
spin dependent hyperfine splitting comes out too small. This has to be
resolved in future calculations. 

The spectrum calculations also allow for the precise determination of
the $b$-quark mass in the $\msbar$-scheme.
The preliminary result is  $m_{b,\msbar}=4.34(7)$ GeV at its
own scale. This result is not affected by the too small hyperfine splitting.

The results for the $D$-spectrum and
the $D_s$ decay constant agree well with experimental results as well
as other lattice calculations, employing a lattice discretisation of
the Dirac action for the charm quark. To obtain reasonable results for
charm quarks is crucial with
respect to the calculation of the form factors for semi-leptonic $B \to
D$ decays.  Recent results for the semi-leptonic 
form factors have been presented as well.

\subsection*{Acknowledgements}
It is a pleasure to thank my collaborators for the ongoing
collaboration on the here presented work. In particular I would like
to thank Sara Collins, Christine Davies and Junko Shigemitsu for their
support in the preparation of this presentation.

\end{document}